\documentclass[preprintnumbers,amsmath,amssymb]{revtex4}

\usepackage{graphicx}
\usepackage{dcolumn}
\usepackage{bm}
\begin{document}
                                                                                                 
\title{The Quantum-Classical comparison of the Arrival Time
Distribution through the Probability Current}
   
\author{\bf{Md. Manirul Ali}\footnote{mani@bose.res.in}$^1$, \bf{A. S. Majumdar}
\footnote{archan@bose.res.in}$^1$ and \bf{Alok Kumar Pan}\footnote{apan@bosemain.boseinst.ac.in}
$^2$} 
            
\address{$^1$ S. N. Bose National Centre for Basic Sciences, Block JD,
Sector III, Salt Lake, Calcutta-700098, India}                                                                                                 
\address{$^2$ Department of Physics, Bose Institute, Calcutta-700009,India}

\maketitle

\vskip 1.5cm

We consider the arrival time distribution defined through the quantum 
probability current for a Gaussian wave packet representing free particles in 
quantum mechanics in order to  explore the issue of the classical limit of arrival 
time. We formulate the classical analogue of the arrival time distribution for an
ensemble of free particles represented by a phase space distribution function 
evolving under the classical Liouville's equation. The classical probability 
current so constructed matches with the quantum probability current in the limit 
of minimum uncertainty. Further, it is possible to show in general that smooth 
transitions from the quantum mechanical probability current and the mean arrival 
time to  their respective classical values are obtained in the limit of large mass 
of the particles.\\

PACS number(s): 03.65.Ta\\


\eject

{\bf 1.INTRODUCTION}\\

It is generally believed that a necessary requirement 
for the universal validity of quantum mechanics is that its results in the 
macroscopic
limit must agree with those of classical mechanics, because the latter
is well verified in the macroscopic domain. However, there exist vexed
problems regarding the connection between classical and quantum mechanics;
the question whether quantum mechanics in the macroscopic limit is completely
equivalent to classical mechanics remains the focal point of diverging view
points. This is poignantly reflected in the various controversies persisting
in the relevant literature \cite{born1}-\cite{dhome}. 
Several naive interpretations of the classical limit of quantum mechanics
based on approaches such as the $\hbar \rightarrow 0$ limit, the large 
quantum number $(N\rightarrow \infty)$ limit, or the Ehrenfest theorem, are
all riddled with well known difficulties \cite{born1}-\cite{pauli}.
Einstein \cite{einstein} and Pauli \cite{pauli} strongly advocated the tenet
that in the macroscopic limit, not only the localised wave functions but
all physically admissible solutions of the Schr${\ddot o}$dinger equation
must lead to predictions equivalent to those obtainable from classical
mechanics. Such comparison between the two mechanics can
be meaningful only within the framework of the ensemble interpretation.
Thus the classical limit problem boils down to probing whether there is
complete equivalence in the macroscopic limit between the empirical
predictions of classical and quantum mechanics with respect to the
properties of the same initial ensemble. This is the spirit which motivates
the present investigation. 

For complete equivalence between classical mechanics and the macroscopic
limit of quantum mechanics the following criterion is necessary. 
In the classical limit all
the measurable properties of a quantum mechanical ensemble corresponding to
any normalizable wave function $\Psi(x,t)$ should be equally reproduced by the
{\it classical phase space formalism} using a distribution function $D(x,p,t)$
utterly determined by the classical phase space description where the 
time-development of $D(x,p,t)$ is in accordance with the classical Liouville's
equation, and  where the initial phase space distribution function for 
the ensemble of particles is taken matching with the initial quantum position 
and momentum distribution.
In the current investigation
we formulate the classical phase space distribution in a way which is 
completely classical unlike the one that is called the 
{\it quantum phase space 
distribution function} such as the Wigner distribution function \cite{wigner}. 
The latter is essentially a quantum entity obtained by directly using the 
expression of the wave function, and is constructed to reproduce the results 
of quantum mechanics, but it does not satisfy the classical 
Liouville's equation. 
So, the Wigner distribution function, not being a positive 
definite quantity in general, does not provide the results of a 
classical 
phase space evolution. In contrast we formulate a  phase space 
distribution function $D(x,p,t)$ that is positive definite and also satisfies 
the classical Liouville's equation. The motivation for this 
work is to study the comparison between quantum mechanical results and those 
obtained from a purely classical phase space description by formulating a 
proper classical counterpart of the quantum ensemble. Here our focus is 
on the {\it arrival time} of the {\it free particles} but one can also 
investigate 
the quantum-classical comparison for other dynamical variables for particles 
in various types of potentials using the same approach.

In recent years there has been an upsurge of interest in understanding
the concept of time of arrival for a quantum particle \cite{leavens}.
In general, the issue of providing  physically meaningful definitions
of experimentally measured times in varied arenas such as tunneling times,
decay times, dwell times, delay times, arrival times has gained 
importance \cite{muga}. In this paper we are specifically concerned with 
the issue of arrival time. In classical mechanics, a particle 
follows a definite trajectory; hence the time at which a particle reaches 
a given location is a well defined concept. On the other hand, in standard 
quantum mechanics, the meaning of arrival time has remained rather 
obscure. Indeed, there exists an extensive literature on the treatment of 
arrival time distribution in quantum mechanics \cite{others}. 
A consistent  approach of formulating a definition for arrival time 
distribution 
is through the quantum probability current \cite{leavens2}.  The quantum 
probability current if defined in an unambiguous manner contains
the spin of a particle, as was pointed out by Holland \cite{holland}.
Recently it has been shown using the explicit example of a Gaussian wave packet
that the spin-dependence of the probability current leads  to the 
spin-dependence 
of the mean arrival time for free particles \cite{ali}. This effect, if 
experimentally observed, should place the probability current approach to mean 
arrival time on a firmer footing. 
A key issue for any definition of time of arrival in quantum 
mechanics is to secure an acceptable classical limit of the arrival
time formulation. We formulate a classical analogue of the 
arrival time distribution for free particles obtained
via the quantum probability current.
Aspects of the quantum-to-classical transition
for the arrival time distribution are then investigated.\\

{\bf 2.ARRIVAL TIME DISTRIBUTION}\\

We begin our analysis with the standard description of the flow of probability 
in quantum mechanics, which is governed by the continuity equation derived 
from  the Schr${\ddot o}$dinger equation given by

\begin{equation}
\frac{\partial}{\partial t}|\Psi({\bf x},t)|^2 + 
{\bf \nabla}.{\bf J}({\bf x},t)=0
\end{equation}

The quantity ${\bf J}({\bf x},t)$=$\frac{i\hbar}{2m}(\Psi {\bf \nabla} 
\Psi^{\ast}-\Psi^{\ast}{\bf \nabla} \Psi)$ defined as the probability current 
density 
corresponds to this flow of probability. We use this current to define the 
arrival 
time distribution for free particles. Interpreting the equation of continuity 
in 
terms of the flow of physical probability, the Born interpretation for the 
squared 
modulus of the wave function and its time derivative suggest that the mean 
arrival 
time of the particles reaching a detector located at ${\bf x=X}$ may be 
written as
\begin{equation}
\bar {\bf \tau}=\frac{\int^{\infty}_{0}|{\bf J}({\bf x}={\bf X}, t)| t  dt}
{\int^{\infty}_{0}|{\bf J}({\bf x}={\bf X}, t)| dt}
\end{equation}
Henceforth, for simplicity we shall restrict ourselves to only one spatial 
dimension.
One should keep in mind  that the definition of the mean arrival time used
in Eq(s).(2) is not a uniquely derivable result within standard quantum
mechanics. However, the Bohmian interpretation \cite{bohmian} of quantum 
mechanics in terms of the causal trajectories of individual particles implies the above 
expression for the mean arrival time in a unique and rigorous way \cite{bohm}.
It should also be noted that in ceratin situations $J(X,t)$ can be negative 
over some time interval provided the initial flux $J(X,t=0)$ is 
negative \cite{bracken}.  In order to account for the back flow effect in
such cases, the decomposition of $J(X,t)$ into right and left moving 
parts could be undertaken. However, our present analysis is carried out
using a simple example that is free from such complications.

The standard Schr${\ddot o}$dinger probability current defined 
through the continuity equation in non-relativistic quantum mechanics,
however, suffers from an inherent ambiguity since the continuity equation
remains satisfied with the addition of any divergence free term to the
current. This feature was exploited to formulate alternative causal
models \cite{deotto}. Finkelstein has analysed the consequent ambiguities
of arrival time distributions \cite{finkelstein}. However, it was shown
earlier by Gurtler and Hestenes that the  problem of non-unique probability 
current doesn't exist
for the relativistic Pauli theory for the electron if the probability current 
is inclusive of a spin-dependent term \cite{hestenes}.
Holland \cite{holland} demonstrated the uniqueness of the conserved probability
current in the non-relativistic limit of the Dirac equation.
This probability current differs from the standard Schr${\ddot o}$dinger probability
current by the presence of a spin-dependent term which persists even in the
non-relativistic limit. It has been further argued that the arrival time
for a free particle computed using the unique probability current should exhibit
an observable spin dependence \cite{ali}. However, for the case of
massive spin-0 particles it has been shown recently by taking the non-relativistic 
limit of  Kemmer equation \cite{kemmer} that the unique probability current is given by the 
Schr${\ddot o}$dinger current, and hence,  the Schr${\ddot o}$dinger current gives 
the unique probability current density or the unique arrival time distribution for 
spin-0 particles \cite{baere}. In the present analysis we restrict our
attention to massive spin-0 particles only.\\

{\bf 3.CLASSICAL-QUANTUM CORRESPONDENCE}\\

Let us now consider a Gaussian wave packet representing a quantum free 
particle moving in 1-D whose initial wave function $\Psi(x,0)$
and its Fourier transform $\Phi(p,0)$ are respectively given by

\begin{equation}
\Psi(x,0)=\frac{1}{(2 \pi {\sigma_0}^2)^{1/4} \sqrt{1+iC}} 
~e^{\left\{{i k x}-\frac{x^2}{4 {\sigma_0}^2(1+iC)}\right\}}
\end{equation}

\begin{equation}
\Phi(p,0)=\left(\frac{2 {\sigma_0}^2}{\pi {\hbar}^2}\right)^{1/4} \
e^{- \left\{\frac{{\sigma_0}^2 (p-{\bar p})^2}{{\hbar}^2} (1+iC)\right\}}
\end{equation}
where the group velocity of the wave packet $u={\hbar k}/m={\bar p}/m$. For
generality we have taken the initial Gaussian wave function $\Psi(x,0)$
which is not a minimum uncertainty
state ($\Delta x \Delta p=(\hbar/2)\sqrt{1+C^2}$ $>$ $\hbar/2$),
but which could represent a squeezed state \cite{robinett}.
The Schr${\ddot o}$dinger time evolved wave function $\Psi(x,t)$, the 
quantum position probability density $\rho_Q(x,t)$ and the 
probability current density  ${J_Q}(x,t)$ at a particular location $x$ 
are then respectively given by

\begin{equation}
 \Psi(x,t)=\frac{1}{(2 \pi {\sigma_0}^2)^{1/4} \sqrt{1+i(C+\frac{\hbar t}
{2 m {\sigma_0}^2})}}\ e^{\it i k (x-\frac{1}{2} u t)}~exp\left\{-\frac{( x- u t)^2}
{4 {\sigma_0}^2 \left[1+i(C+\frac{\hbar t}{2 m {\sigma_0}^2}) \right]}\right\}
\end{equation}
                                                                                                 
\begin{equation}
 \rho_Q(x,t)=|\Psi(x,t)|^2=\frac{1}{(2 \pi {\sigma_0}^2)^{1/2}
\sqrt{1+(C+\frac{\hbar t}{2 m {\sigma_0}^2})^2}}~ exp\left\{-\frac{( x- u t)^2}
{2 {\sigma_0}^2 \left[1+(C+\frac{\hbar t}{2 m {\sigma_0}^2})^2 \right]}\right\}
\end{equation}
                                                                                                 
\begin{equation}
{J_Q}(x,t)=\rho_Q(x,t)~\left\{u +\frac{\hbar (C+\frac{\hbar t}{2 m {\sigma_0}^2}) (x-ut)}
{2 m {\sigma_0}^2 \left[1+(C+\frac{\hbar t}{2 m {\sigma_0}^2})^2 \right]} \right\}
\end{equation}

In order to elucidate the classical counterpart of the quantum probability 
current, we now construct a classical formulation of arrival time for an ensemble of 
free particles. We take the initial phase space distribution function for the ensemble of 
particles as a product of two Gaussian functions matching with the initial quantum 
position and momentum distributions from Eq(s).(3) and (4) as

\begin{equation}
 {D_0}(x_0,p_0,0)=|\Psi(x_0,0)|^2 \hskip 0.1cm |\Phi(p_0,0)|^2
=\frac{1}{\pi \hbar \sqrt{1+C^2}} ~exp\left\{ -\frac{x_0^2}{2 {\sigma_0}^2 (1+C^2)}
-\frac{2 {\sigma_0}^2 (p_0 - {\bar p})^2}{{\hbar}^2} \right\}
\end{equation}

\noindent
where the variables $x_0$ and $p_0$ are the initial positions and momenta of the
particles. Note that our approach to compare the quantum and classical predictions is
not contingent to any particular initial form of the wave function. The key 
point of this scheme is to choose the initial classical ensemble in such a
way that it reproduces the initial quantum position and momentum distributions.
Classically of course there are other choices for $D_0(x_0,p_0,0)$. But in quantum
mechanics, due to the uncertainty principle, given a wave function $\psi(x,t)$,
the momentum space wave function $\phi(p,t)$ is automatically fixed by the 
Fourier transform of $\psi(x,t)$. In this way the position probability density
$|\psi|^2$ and the momentum probability density $|\phi|^2$ are {\it correlated} 
in quantum mechanics. There is no such restriction for the position and momentum 
densities in classical statistical mechanics. But it is quite reasonable to
take the initial classical phase space distribution exactly matching with the
initial quantum position and momentum probability densities in order to compare
the results obtained from the dynamical evolutions of classical and quantum 
mechanics. This is precisely the motivation to take the initial phase space 
distribution $D_0(x_0,p_0,0)$ in a way given by Eq(s).(8). 

Now to obtain the
time evolved density function $D(x,p,t)$ we focus on the classical dynamics
of freely moving particles. The Hamiltonian is $H=p^2/2m$ and the Hamilton's
equations are $x=pt/m +x_0$ and $p=p_0$ where the variables $x_0$ and $p_0$ are the
initial position and momentum of the particle which are respectively given by
$x_0=x-pt/m$ and $p_0=p$. Substituting these values of $x_0$ and $p_0$ in the
expression of $D_0(x_0,p_0,0)$ we obtain the time evolved distribution
function $D(x,p,t)$. This is because here we are considering the free evolution
of an ensemble of particles whose initial positions ($x_0$) and momenta ($p_0$)
are distributed according to the initial density function $D_0(x_0,p_0,0)$. 
The time evolved phase space distribution is then given by

\begin{equation}
D(x,p,t)=\frac{1}{\pi \hbar \sqrt{1+C^2}}~exp\left\{ -\frac{(x-\frac{p t}{m})^2}
{2 {\sigma_0}^2 (1+C^2)}-\frac{2 {\sigma_0}^2 (p-{\bar p})^2}{{\hbar}^2} \right\}
\end{equation}

At this stage it is instructive to write down the the Wigner distribution function which is
calculated \cite{robinett} from the time evolved wave function ($\Psi(x,t)$), 
and is given by 

\begin{equation}
D_W(x,p,t)=\frac{1}{\pi \hbar}\int_{-\infty}^{\infty}\Psi^{\ast}(x+y,t) \Psi(x-y,t)
~~exp\{2ipy/\hbar\} dy
\end{equation}

By substituting the value of $\Psi(x+y,t)$ and $\Psi(x-y,t)$ using Eq(s).(5) we obtain

\begin{equation}
 D_W(x,p,t)=\frac{1}{\pi \hbar} exp\left\{-\frac{2(p-{\bar p})^2 {\sigma_0}^2}{\hbar^2}\right\}
~exp\left\{-\frac{[x-pt/m-2C(p-{\bar p}){\sigma_0}^2/\hbar]^2}{2{\sigma_0}^2}\right\}
\end{equation}

Note here that the Wigner function $D_W(x,p,t)$ is not identical with our 
classical phase 
space distribution $D(x,p,t)$ where in the spirit of a completely classical
description we have not included any position-momentum correlation.

We consider a classical statistical ensemble of particles defined by the phase
space density 
function $D(x,p,t)$ in {\it one dimension}. Then the position and momentum 
distribution functions 
are respectively $\rho_C (x,t)=\int D(x,p,t) dp $ and $\rho_C (p,t)=\int D(x,p,t) dx$. 
The classical position probability distribution for this ensemble is given by

\begin{equation}
 \rho_C (x,t)=\int D(x,p,t) dp=\frac{1}{(2 \pi {\sigma_0}^2)^{1/2} 
\sqrt{1+C^2 +\frac{\hbar^2 t^2}{4 m^2 {\sigma_0}^4}}}
~exp\left\{- \frac{(x-u t)^2}{2 {\sigma_0}^2 (1+C^2 +\frac{\hbar^2 t^2}{4 m^2 
{\sigma_0}^4})} \right\}
\end{equation}

\begin{figure*}
\includegraphics{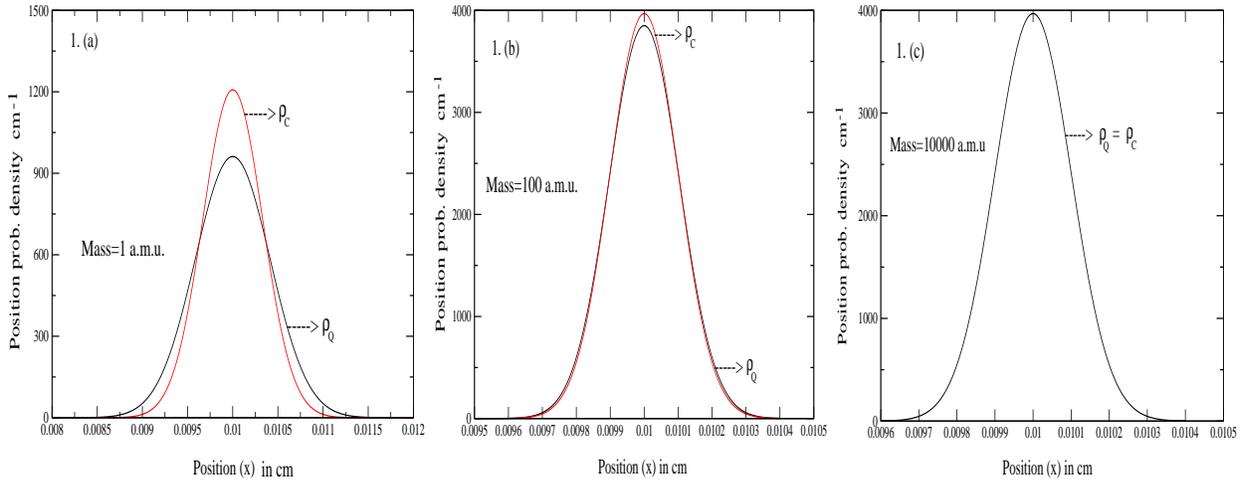}
\caption{\label{fig:wide} The position probability densities ${\rho}_Q(x,t)$
and ${\rho}_C(x,t)$ are plotted for varying mass of the
particles (in atomic mass units) with ${\sigma_0}=10^{-5}$ cm,
$u=10^3$ cm/sec, C=10 and t=$10^{-5}$ sec.}
\end{figure*}

All the density functions are assumed to be normalized and $D(x,p,t)$ 
satisfies the 
classical Liouville's equation\cite{ssg} given by 
\begin{equation}
\frac{\partial D(x,p,t)}{\partial t}+ {\dot x} \frac{\partial D(x,p,t)}{\partial x}
+ {\dot p} \frac{\partial D(x,p,t)}{\partial p} =0
\end{equation}
Since for free particles ${\dot p}=0$ and ${\dot x}=p/m$ we have 
\begin{equation}
\frac{\partial D(x,p,t)}{\partial t}+ \frac{p}{m} \frac{\partial D(x,p,t)}{\partial x}=0
\end{equation}
Integrating the above equation with respect to $p$ one gets 
\begin{equation}
\frac{\partial {\rho_C}(x,t)}{\partial t}+ \frac{\partial}{\partial x}
\left[\frac{1}{m} {\bar p}(x,t) {\rho_C}(x,t) \right]=0
\end{equation}
where ${\bar p}={\int p D(x,p,t) dp}/{\int D(x,p,t) dp}$ is the ensemble 
average of the
momentum. Defining ${\bar v}(x,t)={{\bar p}(x,t)}/m$ as the average velocity, 
one obtains
\begin{equation}
\frac{\partial {\rho_C}(x,t)}{\partial t}+ \frac{\partial}{\partial x}
{J_C}(x,t)=0
\end{equation}
where ${J_C}(x,t)$ and ${\bar v}(x,t)$ represent the mean motion of the 
continuum matter 
at $(x,t)$. Eq(s).(16) is the equation of continuity for the continuous density 
function 
$\rho_C (x,t)$ of a statistical ensemble of particles. The expression for the 
classical probability 
current density is given by
\begin{equation}
{J_C}(x,t)=\frac{1}{m} {\int p D(x,p,t) dp}
\end{equation}
and is related to the mean velocity by ${J_C}(x,t)=\rho_C (x,t){\bar v}(x,t)$.

Now substituting the expression for the time evolved phase space distribution 
function $D(x,p,t)$ from Eq(s).(9) in Eq(s).(17) we get the expression for the current density or the 
arrival time 
distribution at a particular detector location x=X for this classical 
ensemble of free
particles given by
\begin{equation}
{J_C}(x,t)=\rho_C (x,t)~ \left\{ u+ \frac{(x-ut)\hbar^2 t}{\left[\hbar^2 t^2 + 
4m^2 {\sigma_0}^4 
(1+C^2)\right]} \right\} 
\end{equation}
If we impose here the minimum uncertainty condition ${\it viz.}$, 
$C=0$ then one can check from Eq(s).(6), (7), (12) and (18) that 
both ${\rho_Q}(x,t)$=${\rho_C}(x,t)$ and ${J_Q}(X,t)={J_C}(X,t)$ hold,
i.e., the classical and quantum probability currents are similar.
Thus, if we take the initial phase space distribution function for 
the classical ensemble of particles as a product of two Gaussian 
functions matching with the initial {\it quantum} position and 
momentum distributions then the classical arrival time distribution 
exactly matches with the quantum one provided the minimum uncertainty 
relation is satisfied. But in general the quantum and classical 
distribution functions are different when the minimum uncertainty 
condition is not satisfied ($C \ne 0$). 

Though ${J_Q}(X,t)$ and ${J_C}(X,t)$ are in general not equal for
$C \ne 0$, the large mass limits of both are the same. This is seen
from Figures 1 and 2 where the probability distributions and the currents
are plotted respectively for different masses. It is apparent that in 
the large mass limit quantum distributions reduce to the classical 
distributions. The mass dependendence in the arrival time distributions 
and also in the position probability densities (for both the quantum and
classical case) arises from the spreading of the wave packet.

\begin{figure*}
\includegraphics{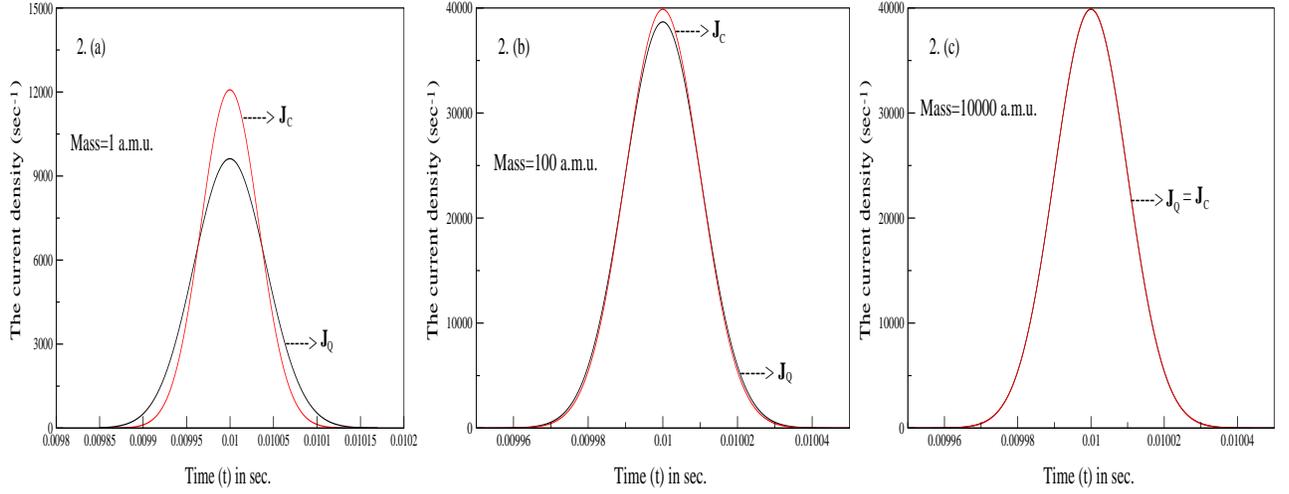}
\caption{\label{fig:wide} The probability current densities $J_Q(x,t)$
and $J_C(x,t)$ are plotted for varying mass of the
particles (in atomic mass units) at a detector location
X=10 cm with ${\sigma_0}=10^{-4}$ cm, $u=10^3$ cm/sec, C=100.}
\end{figure*}

We now compute the mean arrival time $\bar {\bf \tau}$ by substituting 
the expressions for the quantum current in Eq(s).(2). [One should
note that though the integral in the numerator of Eq(s).(2) formally
diverges logarithmically, several techniques have been employed in
the literature \cite{hahne} ensuring rapid fall off for the probability
distributions asymptotically, so that convergent results are obtained
for the integrated arrival time.
Here we have employed a simple strategy of taking  a cut-off ($t=T$) in the 
upper limit
of the time integral with $T=(X+3 \sigma_T )/u$ 
where $\sigma_T$ is the width of the wave packet at time $T$. 
In other words, our computations of the arrival time are valid up to
the $3 \sigma$ level of spread in the wave function.] It is instructive 
to examine the variation of mean arrival time with the different parameters 
of the wave packet. In Figure~3 we have plotted the variation of 
$\bar {\bf \tau}$ with mass at
different detector locations, keeping the group velocity $u$ and initial width 
$\sigma_0$ fixed. One sees that the mean arrival time 
calulated by using the quantum cur÷Cnt $J_Q(X,t)$ as the arrival time 
distribution asymptotically approaches the classical result in the limit
of large mass. 

\begin{figure*}
\includegraphics{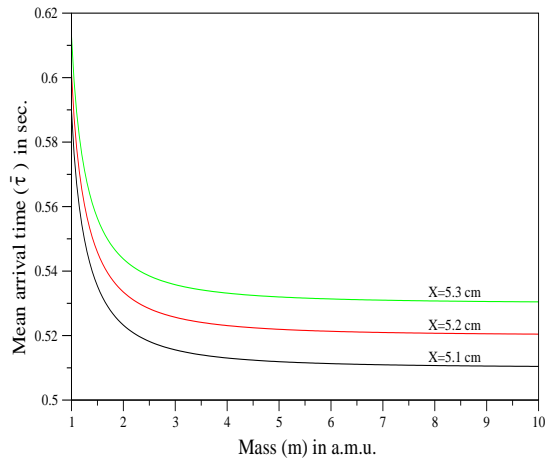}
\caption{\label{fig:wide} The mean arrival time $\bar {\bf \tau}$
is plotted against the mass of the particles (in atomic mass unit) at
different detector locations $X=5.1$ cm, $X=5.2$ cm, and $X=5.3$ cm.
for C=10, ${\sigma_0}=0.0001$ cm, $u=10$ cm/sec.}
\end{figure*}

{\bf 4.SUMMARY AND OUTLOOK}\\

To summarize, in this paper we have investigated the quantum-to-classical 
transition of the mean arrival time defined through the probability current. 
We have formulated the classical  arrival time distribution from the phase 
space distribution for a classical ensemble of particles.  
The expression for classical probability current constructed by us 
matches exactly with the quantum probability current in the limit of minimum 
uncertainty. We note that the uncertainty condition is not a stringent 
requirement for the case of the initial classical distribution. Thus the 
classical arrival time distribution $J_C(X,t)$ will in general be different 
from the quantum distribution $J_Q(X,t)$ if we do not impose the minimum 
uncertainty restriction on the initial distribution. This issue needs to be 
explored further in order to have a deeper understanding of the 
quantum-classical comparison of arrival time. However, in the present example that we have
constructed, the quantum results for the probability current and through it
the arrival time distribution, approaches the classical result in the
large mass limit. A number of schemes \cite{others,bohm,finkelstein,others1}
have been suggested in the literature for calculating
the arrival time distribution such as those based on axiomatic approaches, 
trajectory models of quantum mechanics, attempts to define and calculate
the arrival time distribution using the consistent histories approach,
and attempts of constructing the time of arrival operator, etc. It might be 
worthwhile to investigate the quantum-classical correspondence of the arrival 
time distribution using these different approaches.
Such studies, if undertaken extensively, are not only expected to throw light 
on the comparitive merits of different arrival time formulations, but could
also be of relevance to the behaviour of mesoscopic systems where a great deal
of experimental activity is presently underway \cite{brandes}.

Classically we know that a point particle with uniform motion will reach a 
particular location at a time which is independent of the mass of the particle 
and depends only on its uniform velocity. In the discussion of classical limit 
of quantum mechanics it is usually assumed that the peak mean~position 
of the wave packet moves according to classical trajectory derived  from the 
Ehrenfest theorem. It could be argued though that 
one should not expect to recover an
individual classical trajectory when one takes the classical limit of quantum
mechanics. Rather, one should expect the probability distributions of quantum
mechanics to become equivalent to the probability distributions of an ensemble
of classical trajectories.
The current investigation is concerned about this particular  approach to 
test the
quantitative equivalence between the classical mechanical prediction and the
prediction obtained in the macroscopic limit of quantum mechanics.
Here, what we see is that the {\it mean~time} of 
arrival of a freely moving quantum particle computed through the probability 
current depends on the mass of the particle even if its group velocity is 
fixed. 
So it turns out that the characteristic of mean time in this framework is 
different 
from that of mean position. The predicted mass dependence of mean 
arrival time
is, in principle, amenable for experimental verification, and is a clear 
signature 
of the probability current approach to time in quantum 
mechanics \cite{leavens2}.

\vskip 0.2cm
{\bf Acknowledgments}

We would like to thank Late S. Sengupta, D. Home, C. R. Leavens and 
G. E. Hahne for useful discussions. AKP and MMA acknowledge the Senior 
Research Fellowships of the CSIR, India. 

\vskip 0.5cm



\end{document}